\documentstyle[prb,aps]{revtex} 
\def\narrowtext{} \tighten 
\twocolumn
\newcommand{\REF}{\mbox{\ref{eq:iml1}--\ref{eq:boundeqn}}}
\begin{document}
\draft
\title{
\begin{minipage}[t]{7.0in}
\scriptsize
\begin{quote}
\raggedleft {\rm cond-mat/9805238},\\
{\it Phys. Rev. B,} {\rm in press}
\end{quote}
\end{minipage}
\medskip
\\On the Theory of Magnets with Competing \\
Double Exchange and
Superexchange Interactions}
\author{D. I. Golosov$^{1,2,}$\thanks{E-mail: golosov@franck.uchicago.edu},
 M. R. Norman$^2$,
 and K. Levin$^1$}
\address{
(1) The James Franck Institute, The University of
Chicago, 5640 S. Ellis Avenue, Chicago, IL 60637 \\
(2) Materials Science Division, Argonne National
Laboratory, 9700 S. Cass Avenue, Argonne, IL 60439}
\address{%
\begin{minipage}[t]{6.0in}
\begin{abstract}
In the CMR materials, ferromagnetic double exchange (DE) presumably
coexists with a direct  
nearest-neighbour antiferromagnetic interaction. 
We construct a single-site mean field theory that explicitly takes
into account the
different nature of carrier-mediated ferromagnetism vs. Heisenberg-like
superexchange. 
We find, in contrast to previous
results in the literature,  that the
competition between these two exchange 
interactions  leads to ferro- or
antiferromagnetic 
order with incomplete saturation of the 
magnetization (or sublattice
magnetization), 
rather than spin canting. 
The associated experimental implications are discussed. 
\typeout{polish abstract}
\end{abstract}
\end{minipage}}

\maketitle
\narrowtext

\section{INTRODUCTION}
\label{sec:intro}

Recently, there has been a renewed interest (motivated by
technological problems of microelectronics) in the properties
of colossal magnetoresistance (CMR) manganese oxides \cite{Ramirez} .  
The CMR behaviour typically corresponds to an intermediate doping
range, when these materials are ferromagnetic.
The latter property is generally attributed to 
 a 
conduction electron-mediated double
exchange (DE) interaction\cite{Zener}.
In addition, there exists
 evidence\cite{Aeppli,Osborn,Hirota} which suggests
the presence of 
antiferromagnetic  superexchange interactions of  comparable scale.
In this paper, we study the behaviour of a classical magnet with competing
double exchange and superexchange interactions, and show, in particular, that
in an isotropic case the spin canting (which was previously 
suggested\cite{DeGennes} to be a
generic outcome of such a competition) can be stabilized only at very
high fields and at very low temperatures.

Previous related calculations\cite{DeGennes} have 
been performed 
for a strongly
anisotropic model in which the inter- and intra-layer direct exchange
constants have  different signs \cite{undoped}.
Here, we assume that direct interactions  have everywhere
the same (antiferromagnetic)  sign and  
magnitude. This is viewed as more appropriate for the ${\rm
La_{1-x}Ca_xMnO_3}$ perovskite family  
away from the $x=0$ endpoint, as well as for the layered manganates 
such as ${\rm La}_{2-2x}  
{\rm Sr}_{1+2x} {\rm Mn}_2 {\rm O}_7$. 
In these
compounds, pairs of stacked ${\rm Mn}$--${\rm O}$ planes form
the bilayers, which are separated by poorly conducting layers of ${\rm
La(Sr)O}$ 
(see Ref. \onlinecite{Tokura}). Since the lengths of the intra- and
inter-layer bonds in a bilayer are roughly the same, the values of
inter-layer hopping coefficient and superexchange constant should be
of the same order of magnitude as their intra-layer counterparts.
The observed  interlayer canting\cite{Hirota} or canted 
correlations\cite{Osborn}, presumably caused by 
superexchange between the two layers of a bilayer complex, can be used to 
suggest  relatively large values of superexchange within the layers
as well.
In view of the
considerable interest in these layered  
systems,  we consider
primarily the  two dimensional (2D) 
lattice. By doing so, we expect to capture the basic magnetic properties of
the layered compounds, while avoiding the cumbersome quantitative
treatment of bilayers.  
We also note that our 
2D results are qualitatively representative  of  the three dimensional
case as well. 

We will see that the fact that the double exchange --
superexchange 
competition occurs at {\em all} lattice bonds 
(i.\ e., in the 3D case, for in-, as well as out-of-plane interactions)
leads to the enhancement of spin fluctuations. Lowest
order (i.e.,  Hartree--Fock like)  treatments are
insufficient in this case. Therefore,   in the present paper
we  introduce a new approach to the problem.

We construct a single-site mean field theory which explicitly takes
into account the main feature of the present problem, namely,
the carrier kinetic energy origin of the double exchange ferromagnetism.
We note that the mean field schemes previously reported in the
literature\cite{DeGennes,Millis1} essentially use 
an effective  Heisenberg-like  ferromagnetic exchange interaction to
describe the double exchange. Such schemes would not adequately
reflect the very different
nature
of the two competing interactions.
It is of interest, then, to see how the results are changed if 
a more proper treatment of the double exchange is carried out.
We begin with  the standard
Hamiltonian\cite{DeGennes,Anderson}, corresponding to an infinite on-site 
Hund's rule coupling: 
\begin{eqnarray}
{\cal H}&=& - \frac{t_0}{2} \sum_{<i,j>}\cos 
\frac{\theta_{ij}}{2}\,\{c^\dagger_{i}
c_{j}+c^\dagger_j c_i\} + \frac{J_{AF}}{S^2} \sum_{<i,j>}
\vec{S}_i\cdot\vec{S}_j \nonumber \\ 
&&-\frac{H}{S} \sum_i S_i^z\,.
\label{eq:ham}
\end{eqnarray}   
Here, the first term is the kinetic energy of the carriers (which are
represented by the fermion operators  $c_{j}$ where $j$ 
is the site index).
The second term corresponds to the nearest-neighbour 
antiferromagnetic ($J_{AF}>0$) exchange interaction between the
classical ($S\gg 1$) core (localized) spins  $\vec{S}_i$,
and the last term is the Zeeman energy of these spins in the external 
field $H$.
The double exchange interaction results in the modulation of carrier
hopping coefficients by 
the factors $\cos
\theta_{ij}=\vec{S}_i \cdot \vec{S_j} /S^2$. Since in the present work
we restrict ourselves to a single-site mean field treatment, we omit
the additional phase factors which would result in Berry
phase effects.  
We use units in which the bare hopping
coefficient $t_0$, $\hbar$, $\,k_B$, $\,\mu_B$, and the lattice spacing 
are all equal to unity.

Our mean field framework is based on the exact solution of the single
site problem, which is outlined in the following section. Details of
the derivation are relegated to Appendix \ref{app:iml}, whereas the
implications for the $T=0$ energetics are briefly discussed in
Appendix \ref{app:zerot}. Section \ref{sec:mfe} is concerned with the
mean field treatment of different magnetic phases of the system, and
the resultant mean field phase diagram is described in Section
\ref{sec:phase}. We conclude with a brief discussion of the experimental
relevance of our findings\cite{We}. 

\section{EXACT SOLUTION OF THE SINGLE-SITE PROBLEM}
\label{sec:single}

The random distribution of localized spins leads in
Eqn. (\ref{eq:ham}) to  a highly disordered electronic hopping
problem. Our mean field treatment is based on evaluating  the energy
cost, $\delta F$, of a fluctuation of a single spin $\vec{S}_1$, embedded in an
effective medium with a uniform average value of $\cos \theta_{ij}$.
As we will see below (Eqns. (\ref{eq:btfm}),(\ref{eq:btafm}), and
(\ref{eq:btcm}) ),  
such a fluctuation gives rise to a difference between the values of hopping
coefficient $b$ 
from the site of the 
fluctuating spin  $\vec{S}_1$ to the neighbouring sites 
and the background hopping $t\,\,\,\,\,$ ($t \neq b$); 
for clarity  these parameters\cite{virtual} are
indicated schematically in Fig. 1. 

The quantities
$b$ and $t$ depend on in a self consistent fashion on
the change, $\delta F_{DE}$, in the free energy, associated with the
local change in hopping matrix   
elements $t\rightarrow b$. Such a local change, originating from a local
spin fluctuation on the site $(0,0)$, gives rise to a
perturbation,
\begin{eqnarray}
V=&-&\frac{1}{2}(b-t)\left\{c_{(0,0)}^\dagger
\left(c_{(0,1)}+c_{(1,0)}+c_{(0,-1)}+ \right. \right.
\nonumber \\
 &+&\left.  \left.c_{(-1,0)}\right) + {\rm h. c.}\right\}\,\,,
\label{eq:perturb}
\end{eqnarray}
in the carrier kinetic energy. This perturbation shifts the energy levels
of individual carriers, thus resulting in a change in the total
kinetic energy  of the band.
This kinetic energy contribution to  $\delta F$, which  can be 
evaluated 
following Refs. \onlinecite{Lifshits52,Lifshitsbook,Lifshits47} (see Appendix
\ref{app:iml}) , is given by  
\begin{eqnarray}
\delta F_{DE}(b,t,T)=&& \int f(\epsilon)
\xi(\epsilon) d \epsilon \label{eq:iml1}
\\
&&+ \theta(b-t) \cdot
(\varphi(z_0)-\varphi(-Dt))\,\,, \nonumber 
\end{eqnarray}
where the spectral shift function $\xi(\epsilon)$ takes values between
$-1$ and $1$, and is given by 
\begin{eqnarray}
\xi (\epsilon) =&& - \frac{1}{\pi} \, {\rm Arg} \left\{ b^2 + (t^2-b^2)
\epsilon \int {\cal P}
\frac{\nu(\eta) d\eta} {\epsilon - \eta}+ \right. \nonumber \\
&& \left. + \pi {\rm i} (t^2-b^2) \epsilon \nu(\epsilon) \right\}\,,
\label{eq:iml2}
\end{eqnarray}
the bound
state energy $z_0<-Dt$ is the root of
\begin{equation}
1+\frac{t^2-b^2}{t^2}
\left\{-1+z\int \frac{\nu(\eta) d\eta} {z -
\eta}\right\}=0\,,
\label{eq:boundeqn}
\end{equation}
$\nu(\epsilon)$ is the density of states, 
\begin{eqnarray}
\varphi(z)&=&-T {\rm ln} \left\{ 1+ \exp\left(\frac{\mu-z}{T}\right)\right\},
\nonumber \\
f(z)&=&\frac{1}{\exp \left(\frac{z-\mu}{T}\right) +1}\,,
\label{eq:defphi}
\end{eqnarray}
and $\mu$ is the chemical potential.
Although we will apply 
Eqns.  (\ref{eq:iml1}--\ref{eq:boundeqn})
only to the case of a 2D square lattice, they remain valid for 
a cubic lattice in 3D, as well as in the 1D case. 
The energy integrations are performed over
the conduction band width,  $-Dt< \epsilon, \eta < Dt$ where $D$ is the
dimensionality of the system.   It should 
be stressed that it is because of the locality of the 
perturbation  (\ref{eq:perturb}), 
which represents a lattice analogue of an $s$-wave
scattering problem, that the quantity  $\delta F$ can be 
evaluated 
{\em exactly}\cite{Lifshits52,Lifshitsbook,Lifshits47,Krein}. 

The second term in Eqn. (\ref{eq:iml1}) is a  contribution of a
bound state that appears in the carrier spectrum for $b>t$ (when the
perturbation may be viewed as a ``potential well''). In 2D, the
binding energy of this state vanishes exponentially\cite{exponent}
 as $b \rightarrow
t$, whereas in 3D it has a threshold behaviour. This bound state is
related to one that causes the formation of magnetic polarons
\cite{Nagaev1967}, but it should be distinguished from the true magnetic
polaron which is an extended object and can not be treated within a
single-site approach.

In order to gain additional intuition about the meaning of
Eqns. (\ref{eq:iml1}--\ref{eq:boundeqn}), it is useful to calculate
the energy cost of a single-spin fluctuation in various phases at
$T=0$. This is discussed in Appendix \ref{app:zerot}. This Appendix
highlights the important differences 
between the double exchange and familiar Heisenberg direct exchange
interactions. 

For the purposes of the present work,
the virtual crystal approximation, based on the parameters shown in
Fig. 1,  is expected to
be appropriate  as long 
as the  carrier concentration is not too small. This is because the
quantities of interest involve integration over  carrier
energies in the metallic phase.
In order  to extend this formulation beyond the single site mean field
scheme, we note that  multi-site spin fluctuations  can also be  
treated  
as local perturbations following Ref. \onlinecite{Lifshitsbook}. This  in
principle allows one to  
study 
systematically
the effects of correlations, by constructing an analogue of an
impurity-concentration expansion. This procedure would also verify
whether the virtual crystal approximation is a good starting point for
studying other ({\em e.\ ~g.}  transport) properties.  

\section{THE MEAN FIELD SCHEME}
\label{sec:mfe}

\noindent {\bf 1. Ferromagnetic phase.} In the ferromagnetic phase at
$T>0$, the net energy cost of a 
single-spin fluctuation is (in $2D$) given by  
\begin{eqnarray}
\delta F_1=&&\delta F_{DE}(b,t,T) + 4 J_{AF} \langle 
\cos
\theta_{12} \rangle_2 - H \cos \alpha_1\,- \nonumber \\
&&- 4 J_{AF} \langle \cos
\theta_{12} \rangle_{12}+ H \langle \cos \alpha_1 \rangle_1\,.
\label{eq:domega1fm}
\end{eqnarray}
Here, $\theta_{12}$ is the angle between the directions of the
fluctuating spin   
$\vec{S_1}$ and any of 
The neighbouring spins, denoted by $\vec{S_2}$ (we assume that spin
fluctuations on different sites are statistically independent), 
and  $\alpha_1$ is the angle
between  $\vec{S_1}$ and the direction of magnetization, $\vec{M}$ 
(see
Fig. \ref{fig:canted}). The angular brackets, $\langle
... \rangle_l$,   
are used to denote the average values over the Boltzmann probability 
distribution of spin
$\vec{S_l}$,
$w_l \propto \exp(-\delta F_l/T)$.  We then find that $\langle 
\cos
\theta_{12} \rangle_2=M \cos \alpha_1$, and\cite{averaging}
\begin{eqnarray}
b^2 &\equiv&  \langle \cos^2 (\theta_{12}/2) \rangle_2=
(1+M \cos \alpha_{1})/2\,,\,\,\,\,\,\, \nonumber \\
t^2 &\equiv& \langle b^2 \rangle_1\,=(1+ M^2)/2\,.
\label{eq:btfm}
\end{eqnarray}
The magnetization has to be determined self-consistently as $M = 
\langle\cos 
\alpha_1\rangle_1$; generally, the latter equation has to be solved 
numerically.

In the ferro- and antiferromagnetic phases, it is useful to construct a 
reference framework with which to compare our results. We define 
$J_{eff}(M)$ which represents an effective $M$-
dependent exchange constant for a  Heisenberg-like magnet. The 
appropriate exchange constant can be deduced by considering small 
spin fluctuations ($|\cos \alpha - M| \ll 1$), which 
correspond to small fluctuations in the hopping matrix elements ( $|t-
b|\ll t$). A perturbation expansion 
of Eqn. (\ref{eq:iml1}) then leads to 
\begin{equation}
\delta F_{DE} (b,t,T) \approx -2 \frac{t-b}{t} \int
\epsilon f(\epsilon) \nu(\epsilon) d \epsilon = 2 (t-b) | E_0 |\,, 
\label{eq:smallper}
\end{equation}
 at leading order\cite{finiteT} in $T/t$, where $E_0$ is the kinetic
energy of the
carriers for $t=1$. 
In the ferromagnetic state, it follows from Eqn. (\ref{eq:smallper}) 
that 
\begin{equation}
J_{eff}^{FM}(M)=J_{AF}-
\frac{1}{8}|E_0| \cdot
\sqrt{\frac{2}{1+M^2}}\,. 
\label{eq:JDET}
\end{equation}

The second term in the above equation represents the  DE
contribution. This term, which is  contained in other
mean field  
schemes 
\cite{Millis1,Sarker},  increases as $M$
decreases. As a consequence,  for  moderately strong 
antiferromagnetic
exchange interactions, when\cite{H0}
\begin{equation}
|E_0|  < 8J_{AF} < \sqrt{2} |E_0|\,,
\label{eq:intermediate}
\end{equation} 
$J_{eff}^{FM}(M)$ changes sign
as $M$ varies from $0$ to $1$. This behaviour has important
consequences: it leads to a lack of saturation in the low temperature
magnetization. Typical results for $M(T)$ are plotted in
Fig. \ref{fig:DEFM} for these moderately strong exchange interactions. Here 
the solid line represents the full mean field calculation (which makes use
of Eqns. (\ref{eq:iml1}--\ref{eq:iml2})), while  the dashed line
corresponds to the effective exchange approximation.  The dotted
line represents the behaviour  of a 
conventional Heisenberg magnet with the same value of Curie temperature, 
and a constant nearest-neighbour exchange integral.

The lack of saturation seen in
Fig. \ref{fig:DEFM} can be understood as follows. 
In the paramagnetic phase, $M=0$ and, by virtue of Eqns. 
(\ref{eq:JDET}--\ref{eq:intermediate}), the effective exchange constant has a
negative (ferromagnetic) sign. As $T$ decreases, the system undergoes
a Curie transition at $T_C \approx
4|J^{FM}_{eff}(0)|/3$. 
Decreasing $T$ further results in a decrease in the
magnitude of  spin fluctuations, {\em i. e.}, in an
increase of $M$. The latter is opposed by a  {\em decrease}  in
$|J_{eff}^{FM}|$, leading to the softening of spin fluctuations.  
As a result, the effective exchange constant ``self-adjusts'' in
such a way that it  never becomes large in comparison with
$T$, and even at low $T$ the behaviour of an effective exchange magnet
is similar to that of a conventional Heisenberg magnet in a
``high-temperature'' regime of $T\sim T_C \sim J$.
In this way, the non-vanishing thermal fluctuations do not allow 
the magnetization to reach  its proper saturation value,
$M_0=1$.  At zero field, the value of magnetization  
as $T \rightarrow 0$ is instead given by\cite{fieldM0}
\begin{equation}
M_0=\sqrt{(E_0/J_{AF})^2/32-1} < 1.
\label{eq:M0}
\end{equation}  
These self consistent changes in $|J_{eff}^{FM}|$ lead to inadequacies of the 
effective exchange approximation at low $T$. As may be seen in  Fig. 
\ref{fig:DEFM},  the behaviour obtained in this approximation  differs 
significantly from that found  using the full calculation of $M(T)$.  
This difference is due to the fact that 
when $J_{eff}^{FM}
\stackrel {<}{\sim} T$ is small (in comparison with the electronic
energy scales),   quadratic terms  (in $(t-b)/t$) 
dominate the physics. The details are discussed in Appendix \ref{app:FMinstab}.
Within the effective exchange approximation, strong
fluctuations of {\em both} angular co-ordinates of each spin persist
at low $T$. By contrast, the full calculation shows that the
fluctuations of the  polar angle freeze out, $\cos \alpha_i
\rightarrow M_0$. Independent fluctuations of the azimuthal angles
of the spins, which persist down to $T \rightarrow 0$, appear to be an
artefact of the single-site mean field treatment. It
is natural to expect that, at least in the classical case of $S \gg
1$, these azimuthal fluctuations also freeze out (albeit at a lower
temperature than the polar ones) resulting in the formation of a
multi-sublattice or  spin-glass-like state with the net magnetization
given approximately by Eqn. (\ref{eq:M0}).

\noindent {\bf 2. Antiferromagnetic phase.}
The N\'{e}el antiferromagnetic state (of the metallic phase) can be 
treated 
similarly. 
We find $\langle \cos \theta_{12} \rangle_2 = - m \cos \alpha_1$,
where $\alpha_1$ is the angle formed by
the spin $\vec{S}_1$ with its average direction and  $m$ 
is sublattice magnetization.  Eqns.
(\ref{eq:btfm}) are replaced by 
\begin{equation}
b^2=(1-m \cos \alpha_1)/2\,,\,\,\,\,\,\, 
t^2=(1-m^2)/2\,.
\label{eq:btafm}
\end{equation}
 Instead of (\ref{eq:JDET}) we obtain
\begin{equation}
J_{eff}^{AFM}(m)=J_{AF}-\frac{1}{8}|E_0|\sqrt{\frac{2}{1-m^2}}\,.
\end{equation}
It is easy to show that the  N\'{e}el ordering  
arises  for 
$J_{AF}> 
2^{-5/2} |E_0|$ in zero field and at $T<T_N \approx 4J_{eff}^{AFM}(0)/3$.  
It always exhibits undersaturation  of the 
sublattice magnetization: at $T\rightarrow 0$,
\begin{equation}
m\rightarrow m_0=\sqrt{1- (E_0 / J_{AF})^2/32}<1\,.
\end{equation}
This undersaturation (which leads to a 
finite bandwidth) may be 
viewed as
consistent with the presumed  metallic state. 

Numerical calculations yield the dependence of $m$ on $T$, which is 
similar to $M(T)$ in the ferromagnetic phase and shows the same
low-temperature features. 

\noindent {\bf 3. Canted phase.}
Our discussion thus far has not included the canted phase first 
proposed by 
De Gennes\cite{DeGennes}. 
In our case, this is a two-sublattice (checkerboard) magnetic phase; 
the sublattice magnetizations have an equal magnitude  $m$
and form  an angle $2 \gamma$ with each other.  In the present model, spin
canting  requires the presence of a  magnetic
field to break the high degeneracy which would otherwise occur.  
This degeneracy is related to the fact that  the energy of
the system depends solely on 
the values of the angles formed by the pairs of neighbouring spins.
All the neighbours of any  spin $\vec{S}_1$ of sublattice I belong
to sublattice II, and are parallel to each other at $T=0$. 
Therefore, the energy of the system does not change as the  spin $\vec{S}_1$
moves along  any cone around their common direction. In the context of
single site  mean field approaches, the same holds at $T>0$ for any
cone around the {\em average} direction of the sublattice II spins.  
Thus, the probability distribution of the spin $\vec{S}_1$ will be
axially-symmetric with respect to the direction of the magnetization of
sublattice II, with which    
the spin $\vec{S}_1$ will therefore be aligned on average  (rather
than with  sublattice I). 
 Thus, in the absence of perturbations
(caused by next-nearest-neighbour exchange, anisotropy effects,
quantum corrections, or small  
external fields) the canted state is destabilized, as a result of the
underlying degeneracy\cite{layered}.   Since it is
site-local \cite{Kagome}, its effects  will persist  as long as 
the energy
scale of a perturbation {\em per individual spin} remains small in
comparison with
the characteristic energy, $k_B T$, of the 
thermal motion of a {\em single} spin.

To characterize the finite field canted state, we use 
the full non-perturbative expression (\ref{eq:iml1}). The mean field
framework of Eqns. (\ref{eq:domega1fm}--\ref{eq:btfm}) has to be
modified to allow for a self-consistent determination of the {\it two}
mean-field variables, $m$ and $\gamma$.
We now obtain two coupled mean field equations, which, as in the
ferromagnetic phase, follow from 
the self-consistent definition of the sublattice magnetization $m$.
For   the 
component
of $\langle \vec{S}_1 \rangle$ parallel to the magnetization of
sublattice I, we obtain  
\begin{equation}
-\sin 2 \gamma \,\langle \sin \alpha_1 \, \cos \beta_1 \rangle_1 + 
\cos 2
\gamma \,\langle \cos \alpha_1 \rangle_1=m\,, 
\label{eq:mfec1} 
\end{equation}
whereas the perpendicular component must vanish,
\begin{equation}
\cos 2 \gamma \, \langle \sin \alpha_1 \, \cos \beta_1 \rangle_1 + 
\sin 2
\gamma \, \langle \cos \alpha_1 \rangle_1 =0\,.
\label{eq:mfec2}
\end{equation}
In writing Eqns. (\ref{eq:mfec1}--\ref{eq:mfec2}) we used a co-ordinate
system with a polar axis parallel to the sublattice II magnetization.
$\alpha_1$ and $\beta_1$ are polar and azimuthal angles of 
the
spin $\vec{S_1}$ in this frame, with  $\beta_1=0$ corresponding
to the spin $\vec{S_1}$ lying within the plane 
containing the two sublattice magnetizations. 

For the net energy of a single-site fluctuation we now obtain, instead
of Eqn. (\ref{eq:domega1fm}),
\begin{eqnarray}
\delta F_1&&=\delta 
F_{DE}(b,t,T) + 4 J_{AF} m \cos \alpha_1-  4 J_{AF} m \langle 
\cos \alpha_1 \rangle_1 \,- \nonumber \\
&&- H(-\sin 
\gamma\,\sin
\alpha_1\,\cos \beta_1+\cos\gamma \cos \alpha_1)+ \nonumber \\
&&+ H(-\sin 
\gamma\,\langle \sin \alpha_1\,\cos \beta_1\rangle_1\,+\cos\gamma
\langle \cos \alpha_1\rangle_1\,)
\end{eqnarray}
whereas the values of the hopping coefficients (see Fig. 1 and
Eqn. (\ref{eq:perturb}))  are given by
\begin{equation}
b^2=(1+m\cos
\alpha_1)/2 \,\,,\,\,\,\,\,\,\,t^2=(1+m^2 \cos 2 \gamma)/2\,.
\label{eq:btcm}
\end{equation}
The low-$T$ canted state is found to be stable for $8 
J_{AF}>|E_0| +  H$. 

We begin with the case of relatively large bandfilling, corresponding
to the undersaturated ferromagnetic behaviour at $H=0$ (see Eqn. 
(\ref{eq:intermediate}) and Fig. \ref{fig:DEFM}).
The  solutions\cite{integral} of
Eqns. (\ref{eq:mfec1}--\ref{eq:mfec2}) for typical parameters   are 
illustrated  in 
Fig. \ref{fig:Mcanted}.  
One can see  that,
as $T \rightarrow 0$ in the canted phase, the sublattice 
magnetization $m$
approaches its proper saturation value $m=1$.  Note that the
ferromagnetic  ($\gamma =0$) 
solution to the mean field equations is
present at $H>0$ as well. In Fig. \ref{fig:Mcanted}, the
corresponding magnetization, $M_{FM}(T)$, is represented by the 
dotted line.  
However, when the canted ($\gamma > 0$) solution exists, it 
corresponds
to a lower value of the free energy. This is obvious from the fact
that the net magnetization in the canted state $M_{CM}(T) \equiv -
\partial F /\partial 
H = m \cos \gamma$ (dashed line in Fig. \ref{fig:Mcanted})
is larger than   $M_{FM}(T)$. The canted solution branches 
from the
ferromagnetic one at  
temperature  $T_1
\sim H $, when 
\begin{equation}
4TM=H \langle \sin^2 \alpha_1 \rangle_1\,;
\label{eq:T1}
\end{equation}
 at this point the
undersaturated ferromagnetic state undergoes a second-order spin-flip
transition into the low temperature canted state\cite{susce}.
One can therefore conclude that {\it undersaturation is 
representative of the generic low-temperature behaviour of a double
exchange --
superexchange magnet}\cite{Chubukov}.

For smaller values of carrier concentration, at $H>0$ we find 
the spin-flop phase\cite{flip} of the undersaturated antiferromagnet, which
evolves into the canted state via a smooth crossover at $T \sim H$, as
shown in Fig. \ref{fig:flop}. The low-temperature region where the canting
angle $\gamma$ rapidly increases with $T$ corresponds to the canted phase.

\section{MEAN FIELD PHASE DIAGRAMS}
\label{sec:phase}

Typical phase diagrams for the DE--superexchange magnet in  
 (a,c) zero 
and (b,d) non-zero field  
are presented  in
Fig. \ref{fig:phase}. For $t_0$ of the order of an eV, our
choice of parameters corresponds to reasonable values of $J_{AF}
\stackrel{<}{\sim}300 {\rm K}$. In zero field (a,c),
the solid line represents the phase boundary
between paramagnetic
(PM) and antiferro- (AFM) or ferromagnetic (FM) metallic phases.
For the values of parameters used in Fig. \ref{fig:phase} (a), the
ordered phases are undersaturated at low $T$. For slightly  
smaller $J_{AF}$ we find a 
critical value of bandfilling,  $x_1$, which divides the saturated,
$x>x_1$, and undersaturated regimes (see Fig.  \ref{fig:phase} (c)).
At low temperatures and small concentrations,
the undersaturated AFM state becomes thermodynamically
unstable ($\partial \mu /\partial x < 0$),  signalling either the
onset of a more
complicated spin arrangement or phase separation (see Appendix
\ref{app:FMinstab}). The dashed line 
in
Fig. \ref{fig:phase} (a,c) corresponds to the anticipated boundary of
this region 
($\partial \mu /\partial x = 0$). We note that the possibility of
phase separation in DE--superexchange  systems has been
suggested both by analytical studies\cite{Nagaevreview} and numerical
simulations\cite{Dagotto}. 

Figs. \ref{fig:phase} (b,d) show that in the
presence of a magnetic field the PM--FM transition is replaced by a
smooth crossover (dotted line).  The spin arrangement  
of the AFM phase becomes non-collinear (flop-phase), and has
the same symmetry properties as the canted phase (CM), 
which becomes stable at lower $T$ (replacing the $H=0$ 
undersaturated FM and AFM phases). The two are separated from 
the PM
and FM region by a second-order phase transition at $T=T_1(x)$, which is 
represented
by the solid line. At
sufficiently small $x$ the latter approaches the $H=0$  N\'{e}el
transition line. The thermodynamic instability line (not shown in
Fig. \ref{fig:phase} (b,d)) 
is only slightly affected by $H$.

\section{DISCUSSION}
\label{sec:conclu}

We expect that our calculations are directly relevant to the quasi-2D
layered materials ${\rm La}_{2-2x}  {\rm Sr}_{1+2x} {\rm Mn}_2 {\rm O}_7$.
The existence of a strong superexchange interaction in this system is
suggested by 

\noindent (i) relatively high values of N\'{e}el temperatures observed at the
$x=1$ endpoint \cite{Marissa}, 

\noindent (ii) intra-layer antiferromagnetic correlations present near $T_C$
(Ref. \onlinecite{Aeppli}),   

\noindent (iii) interlayer (within the same bilayer) canting found at
low temperatures\cite{Hirota} (see also Ref. \onlinecite{layered})
and interlayer canted
correlations\cite{Osborn} present above $T_C$.

The latter point is associated with the structure of the quasi-2D manganates,
which was discussed in the Introduction.

The verification of the undersaturated behaviour ot low $T$ remains
an open question. 
It is not clear whether the materials ${\rm La}_{2-2x}  
{\rm Sr}_{1+2x} {\rm Mn}_2 {\rm O}_7$,   with $x=0.4$,  
lie 
within the region where the system exhibits
undersaturated ferromagnetic behaviour at low $T$, or outside 
of this region (in the latter case, we still expect thermal
fluctuations to be stronger than 
in a Heisenberg magnet, due to the presense of superexchange). 
Some measurements of the absolute value of magnetization\cite{difference} in  
$x=0.4$
samples indicate undersaturation \cite{Rietveld,Raveau}, while 
others
do not \cite{Moritomo}.
 
We suggest that magnetic properties of the samples exhibiting 
undersaturation should
be studied in the high-field, low-temperature regime of $T \stackrel 
{<}{\sim}H$. Our results (see Eqn. (\ref{eq:T1})) indicate that the
intra-layer canted spin ordering should be stabilized in this region.   
Another important prediction of our theory is the unusual dependence
of the effective ferromagnetic exchange constant on magnetization and
hence on temperature (Eqn. (\ref{eq:JDET})).
While we did not study spin waves in the undersaturated low-temperature
phase, it is clear that in such a situation the usual relationship
between the low-$T$ value of spin stiffness $D_0$ and the Curie temperature
($D_0 \propto T_C$) is no longer valid. This might
help explain the recent experimental findings in perovskite manganates
\cite{Fernandez}.
We propose that the
magnetization dependence of the effective exchange constant 
(available through spin wave measurements) should be studied in more 
detail in both 3D and 2D systems. 

It should be noted that the presence of undersaturation in ferro- and
antiferromagnetic 
phases may well signal that in reality the system favours more
complicated ({\em e.g.} spin glass-like, cf. Ref. \onlinecite{Moritomo})
spin ordering, that cannot be addressed within a single-site mean
field theory. The fact that there have been no observations of  
an {\em intra-layer} spin canting  in the layered
compounds at $T \gg H$ is consistent with our results.

\acknowledgements

This work has benefited from enlightening discussions that we had
with many theorists and experimentalists, especially  A. G. Abanov,
A. Auerbach, 
A. V. Chubukov, M. I. Kaganov, M. Medarde, J. F. Mitchell, 
R. Osborn, R. M. Osgood,
T. F. Rosenbaum, and A. E. 
Ruckenstein. We
acknowledge the support of a Univ. of Chicago/Argonne National
Lab. collaborative 
Grant, U. S. DOE, Basic Energy Sciences, Contract
No. W-31-109-ENG-38, and the MRSEC program of the NSF under 
award \#
DMR 9400379.

%\verb+\appendix+
\appendix
\section{Derivation of Eqns. (\REF).}
\label{app:iml}

We begin by re-writing the local perturbation (\ref{eq:perturb}) as
\begin{equation}
V=- (b-t)(a_1^\dagger a_1-a_2^\dagger a_2)
\label{eq:newper}
\end{equation}
in terms of the fermion operators,
\begin{equation}
a_{1,2}=\frac{1}{\sqrt{2}} c_{(0,0)} \pm \frac{1}{2 \sqrt{2}} \left(
c_{(1,0)}+c_{(0,1)}+c_{(-1,0)}+c_{(0,-1)}\right)\,, 
\label{eq:newstates}
\end{equation}
which anti-commute with each other. Perturbations of this form can be
treated exactly by following  I. M. Lifshits' theory of local
perturbations \cite{Lifshits52,Lifshitsbook}. Here we will use mainly
the Green's functions (resolvent operators) approach of
Refs. \onlinecite{Lifshitsbook,Krein}. We will, without loss
of generality, consider the 2D case.

Perturbation (\ref{eq:newper}) results in a change of the net free
energy of the carriers, which can be evaluated as
\begin{eqnarray}
\delta F = \delta \Omega = &&- T \,{\rm Tr} \left\{ {\rm ln} \left[ 1+ \exp
\left( \frac{\mu - {\cal H}_{vc} -V}{T} \right) \right] -
\right.  \nonumber \\
&& \left.-{\rm ln} \left[ 1+ \exp
\left( \frac{\mu - {\cal H}_{vc}}{T} \right) \right] \right\} =
\nonumber \\
=&&\int_{-\infty}^\infty \varphi(\epsilon) \left( \tilde{\nu}_{tot} (\epsilon)
- \nu_{tot}(\epsilon) \right) d \epsilon \,. 
\label{eq:krein}
\end{eqnarray}
Here, $\varphi(\epsilon)$ is defined by Eqn. (\ref{eq:defphi}),
$\tilde{\nu}_{tot}(\epsilon)$ is the total (for the entire system)
carrier density of states in the 
presence of 
perturbation (\ref{eq:newper}), and
\begin{equation}
\nu_{tot}(\epsilon)=N \nu(\epsilon)=\frac{4N}{\pi^2} \cdot
\frac{1}{2t+\mid\epsilon\mid} {\cal K}
\left(\frac{2t-\mid\epsilon\mid}{2t+\mid\epsilon\mid}\right)\,\,,
\label{eq:dos}
\end{equation}
(where ${\cal K}(x)$ is the complete elliptic integral and N is the
number of lattice sites) is the total density of
states corresponding to the unperturbed 
virtual-crystal band Hamiltonian, 
\begin{equation}
{\cal H}_{vc} = - \frac{t}{2} \sum_{\langle i,j \rangle}(c^\dagger_i
c_j + c^\dagger_j c_i)\,,
\end{equation} 
with the spectrum, $\epsilon(\vec{q})=-t(\cos q_x + \cos q_y)$. The
perturbed density of states, $\tilde{\nu}_{tot}(\epsilon)$, may include
$\delta$-function peaks corresponding to the discrete levels
which split off downwards from the bottom of the band. As we shall see
below, only one discrete level may appear in the present problem\cite{top},
and after integration by parts the free energy change (\ref{eq:krein})
can be re-written 
\cite{Lifshits52,Krein} in
the form of Eqn. (\ref{eq:iml1}) (where the first term on the r.\ h.\
s. accounts for the contribution of the continuous part of the
spectrum). We note that  Eqn. (\ref{eq:iml1}) is an example of a Krein
trace formula. The quantity  
\begin{equation}
\xi(\epsilon) = - \int_{-\infty}^\epsilon \left( \tilde{\nu}_{tot}(\eta) -
\nu_{tot}(\eta)\right) d \eta
\end{equation}
is called the {\it spectral shift function} because of its
relationship to the perturbation-induced shifts of the energy levels
in the case of a discrete (or discretized) unperturbed spectrum
\cite{Lifshits52}. It can be evaluated as
\begin{equation}
\xi(\epsilon) = \frac{1}{\pi} {\rm Im}\, {\rm Tr} \left\{ {\rm ln}
G(\epsilon - {\rm i}0) -  {\rm ln}G_0(\epsilon - {\rm i}0) \right\}\,,
\label{eq:xieval}
\end{equation}
where the operators $G_0(\epsilon) = (\epsilon \cdot \hat{1} - {\cal
H}_{vc})^{-1}$ and 
\begin{equation}
G(\epsilon) = \left( G_0^{-1}(\epsilon) - V\right)^{-1} =
\left(\hat{1} - G_0(\epsilon) V \right)^{-1} G_0(\epsilon)
\label{eq:geval}
\end{equation}
are the Green's functions for the unperturbed and perturbed
Hamiltonians, respectively, and $\hat{1}$ is the identity operator.
Eqn. (\ref{eq:xieval}) yields $d \xi /d \epsilon = - {\rm Im \,} {\rm
Tr} \{G(\epsilon - {\rm i}0) - G_0(\epsilon - {\rm i}0)\}/\pi$. In turn,
\begin{eqnarray}
&&{\rm Tr} (G-G_0)={\rm Tr}\left\{ (\hat{1}-G_0 V)^{-1} G_0 V G_0
\right\} = \nonumber\\ 
&&={\rm Tr} \left\{ (\hat{1}-G_0 V)^{-1} G_0^2 V \right\} = 
\nonumber \\
&&=\frac{d}{d \epsilon} \,{\rm Tr}\,{\rm ln}(\hat{1}-G_0 V) =
\frac{d}{d \epsilon}\, {\rm ln}\, {\rm Det} (\hat{1} - G_0 V),
\end{eqnarray}   
where we used the fact that the operators $(\hat{1} - G_0 V)^{-1}$ and
$G_0 V$ commute with each other. Therefore \cite{Lifshitsbook,Krein},
\begin{equation}
\xi(\epsilon) = - \frac{1} {\pi}\, {\rm Arg}\,{\rm Det}\left\{ \hat{1} -
G_0(\epsilon - {\rm i}0) V \right\}\,.
\label{eq:xifinal}
\end{equation}
Since, according to Eqn.  (\ref{eq:newper}), the perturbation $V$ is
nothing but the sum of two projection operators, it
 is convenient to evaluate the r.\ h.\ s. of Eqn. (\ref{eq:xifinal})
in a basis which includes the states $|1 \rangle$, $|2 \rangle$, 
annihilated by the operators $a_{1,2}$ (see
Eqn. (\ref{eq:newstates})). 
In this basis, the determinant reduces to that of a $2 \times 2$
matrix, and one obtains
\begin{eqnarray}
&&\xi=-\frac{1}{\pi} \,{\rm Arg} \left\{  1 + (b-t) (I_{11}-I_{22}) - (b-t)^2
\times \right. \nonumber \\
&&\left. \times\left[ {\rm Det}
(I_{ij})- \pi^2\, {\rm Det}(C_{ij})\right]+ \pi {\rm i} \left[
(b-t)(C_{11}-C_{22})+ \right. \right. \nonumber \\
&& \left. \left. +(b-t)^2(C_{12}I_{21}+C_{21}
I_{12}-C_{11}I_{22} -C_{22}I_{11}) \right] \right\}\,.
\label{eq:xiinter}
\end{eqnarray}
Here the quantities $I_{ij}$ and $C_{ij}/\pi$ with $i,j = 1,2$
denote, respectively, the 
real and imaginary parts of the matrix elements $\langle i
|G_0(\epsilon - {\rm i}0)| j \rangle$. Explicitly, we find
\begin{eqnarray}
C_{11}(\epsilon) &=& \frac{1}{2}\nu(\epsilon) \left(1 -
\frac{\epsilon}{t} \right)^2\,, \,\,\,\, \\
C_{12}(\epsilon) &=& C_{21}(\epsilon)= \frac{1}{2} \nu(\epsilon) \left(
1 - \frac{\epsilon^2}{t^2} \right)\,, \nonumber \\
C_{22}(\epsilon)&=& \frac{1}{2} \nu(\epsilon) \left( 1
+\frac{\epsilon}{t} \right)^2\,,\,\,\,\,\,I_{ij}(\epsilon)=
\int_{-2t}^{2t} {\cal P} \frac{C_{ij}(\eta) d \eta}{\epsilon -
\eta}\,. \nonumber
\end{eqnarray}
We then obtain\cite{dos}
\begin{equation}
{\rm Det} (C_{ij}) = 0\,,\,\,\,\,{\rm Det} (I_{ij})= - \frac{1}{t^2}
+ \frac{\epsilon}{t^2} \int_{-2t}^{2t} {\cal P} \frac{\nu(\eta) d
\eta}{\epsilon - \eta}\,, 
\end{equation}
etc., and finally, Eqn. (\ref{eq:xiinter}) takes form of Eqn. (\ref{eq:iml2}).
The latter can be conveniently re-written as
\begin{equation}
\pi\, {\rm cot}\{ \pi \xi (\epsilon)\} = - \frac{1}{\epsilon \nu(\epsilon)}
\frac{b^2}{t^2-b^2} - \frac{1}{\nu(\epsilon)} \int {\cal P}
\frac{\nu(\eta) d\eta} {\epsilon - \eta}\,, 
\label{eq:iml2old}
\end{equation}
where the branch of ${\rm arc \, cot}$ should be selected in a way which
respects both the continuity of $\xi(\epsilon)$ \cite{Krein} and the fact
that $\xi(\epsilon) \equiv 0$ for $b=t$. 

We note that the r.\ h.\ s. of the Eqn. (\ref{eq:iml2old}) has the form
$F(\epsilon)/\nu(\epsilon)$, diverging as $\nu(\epsilon) \rightarrow
0$. Therefore its possible values below the bottom of the band are
$\mp \infty$,
corresponding either to $\xi(\epsilon)=-1$ or to $\xi(\epsilon)=0$. The case of
$\xi(\epsilon)=-1$ corresponds to the values of
$\epsilon$ between the 
bottom of the band and the bound state when the
latter is present, while for $\epsilon$ smaller than all the
eigenvalues of ${\cal H}$ (continuous and discrete alike) the spectral
shift function vanishes, $\xi(\epsilon)=0$.
The change between these two
values, which corresponds to the bound state, can occur only at
$\epsilon = z_0$, where $z_0$ satisfies the equation $F(z)=0$. The
latter condition yields Eqn. (\ref{eq:boundeqn}). Although it appears rather
intuitive, this  consideration of the bound-state problem can be
substantiated by a direct calculation along the lines of
Ref. \onlinecite{Lifshits47}. 

Interestingly, in  Ref. \onlinecite{Lifshits52}  the notion of a
finite trace of certain operators in an infinite-dimensional Hilbert space (see
Eqn. (\ref{eq:krein})) was essentially introduced for
the first time. Mathematical studies of 
related issues were initiated by M. G. Krein\cite{Krein}, and since
then the Krein trace formulae remain an active research topic of
functional analysis. 

\section{SATURATED PHASES AT T=0}
\label{app:zerot}

In this Appendix, we present  some numerical and analytical
results related to the phases which saturate at low $T$ (i.\ e., 
the phases with values of
magnetization or sublattice magnetization approaching unity at $T
\rightarrow 0$). This being the simplest application of Eqns.
(\ref{eq:iml1}--\ref{eq:boundeqn}), it provides  insight into the
meaning of these equations which are crucial for the present paper.
The undersaturated phases at low $T$ will be considered in the
Appendix \ref{app:FMinstab}.

We will assume that in the ground state,
all the pairs of neighbouring spins form the same angle, $2 \gamma$,
and that each spin forms an angle $\gamma$ with the $z$ axis
(thus, $\gamma = 0 $ corresponds to the ferromagnetic phase, and $\gamma
> 0$ -- to the two-sublattice canted state of De Gennes).
Then the  energy of the system at $T=0$ can be written as \cite{DeGennes}
\begin{equation}
F^{(0)}/N = - \mid E_0 \mid \cos \gamma + 2 J_{AF} \cos 2 \gamma -
H \cos \gamma,
\label{eq:omega0}
\end{equation}
Here, $E_0$ is the energy of electrons for $t=1$.
Let us now consider a single-spin perturbation of the ground state
corresponding to the change of the polar angle value of a spin $\vec{S_1}$
from $\gamma$ to $\alpha$. The energy difference between this
configuration and the ground state is given by
\begin{eqnarray}
&&\delta F^{(0)}(\alpha)=\delta F_{DE}(b(\alpha),t,0)+ 
\nonumber \\
&&+4 J_{AF}
\left( \cos(\alpha + \gamma) - \cos 2 \gamma \right) - H(\cos \alpha -
\cos \gamma)\,,
\label{eq:dF0}
\end{eqnarray} 
where $\delta F_{DE}(b(\alpha),t,0)$ is given by Eqns.
(\ref{eq:iml1}--\ref{eq:boundeqn}) with $t=\cos \gamma$ and $b=
\cos\{(\alpha + \gamma)/2\}$. 
By minimizing $E^{(0)}$ with respect to
$\cos \gamma$, we find \cite{DeGennes} that the ferromagnetic phase is
stable at $\mid E_0 \mid + H \geq 8 J_{AF}$. In this case one can use
Eqn. (\ref{eq:smallper}) to obtain the value of $\delta F^{(0)}$ for
$\alpha \ll 1$:
\begin{equation}
\delta F^{(0)} (\alpha)= -(4 J_{eff}^{(0)}- H)(1- \cos
\alpha)\,, 
\label{eq:eeaT0}
\end{equation} 
where the effective exchange constant (cf. Section \ref{sec:mfe}) is
given by $J_{eff}^{(0)}=J_{AF}-|E_0|/8$. 
In the case of pure double exchange, $J_{AF}=0$, and for
sufficiently large carrier concentration $x \stackrel {>}{\sim} 0.1$,
the effective exchange approximation is in fact adequate even for
large values of $\alpha$. Numerically, the difference between Eqns.
(\ref{eq:dF0}) and (\ref{eq:eeaT0}) at $\alpha = \pi$ does not exceed
15--20 \%. This relative difference (which reflects the different
physics of the
double exchange and Heisenberg exchange) becomes more pronounced at
large $J_{AF} \sim |E_0|/8$ (see Fig. \ref{fig:domegafm}).  
At smaller concentrations, $x < 0.27$ in 2D, 
and at sufficiently
large values of $J_{AF}$, we find $\delta F^{(0)}(\pi) < 0$
(dotted line in Fig.  \ref{fig:domegafm}). This
means that the energy of the system can be lowered by flipping a
single spin, and the  ferromagnetic state becomes metastable.

For larger values of $J_{AF}$, corresponding to
$J_{AF}>(|E_0| + H)/8$, the canted state with 
\begin{equation}
\cos \gamma = \frac{\mid E_0 \mid + H}{8 J_{AF}}\,
\label{eq:gammaT0}
\end{equation}
emerges at $T=0$, $H>0$ (see Sect. \ref{sec:mfe} regarding the latter
condition). In this case, the energy of small single-spin perturbations
is quadratic in $|b-t|/t \ll 1$,
\begin{eqnarray}
&&\delta F^{(0)}(\alpha) \approx \left\{\left(\mid E_0 \mid + \frac{ 16 H
J_{AF}^2}{64 J_{AF}^2-(\mid E_0 \mid+H)^2}\right) \cos \gamma
 - \right. \nonumber \\
&& - \left.
\int_{-2}^{\mu}\left( \int_{-2}^{2} {\cal P}\frac
{\nu(\eta) d 
\eta}{\eta - \epsilon}\right) \epsilon^2 \nu(\epsilon) d \epsilon
\right\} (\alpha - \gamma)^2\, {\rm tan}^2 \gamma\,.
\label{eq:cantedinstab}
\end{eqnarray}  
Note that the effective exchange approximation, which is based on the
first-order  (in $(b-t)/t$) perturbation theory result
(\ref{eq:smallper}), is inapplicable.

The typical results for $\delta F^{(0)} (\alpha)$ in the canted state
are shown in Fig. \ref{fig:domegacanted} (left panel), 
where the dashed line represents the contribution of the band (first) 
term on the r.\ h.\ s. of Eqn. (\ref{eq:iml1}); one can see that the 
bound state noticeably lowers energies of fluctuations with 
$\alpha \approx 2 \pi - \gamma$. 

The origins of instabilities of the canted state which appear in our 
single-site treatment are 
illustrated in the right panel of Fig. \ref{fig:domegacanted}, 
where  functions $\delta
F^{(0)} (\alpha)$ at different band fillings $x$ for $J_{AF}=0.06$,
$H=0.01$ are plotted. We see that as one lowers the bandfilling from
$x=0.4$ to $x=0.25$,  $\delta F^{(0)}(\pi - \gamma)$ becomes
negative, so that the total energy can be lowered by flipping a single
spin of sublattice I to the direction antiparallel to that of
sublattice II spins, and the canted state is metastable.
As one further lowers concentration
to $x=0.15$, the sign of $\partial^2 \delta
F^{(0)} (\alpha) /\partial \alpha^2 $ at $\alpha =\gamma$
changes, signalling the instability of the canted phase.
Indeed, since in 2D the principal-value integral on the r.\ h.\ s. of
Eqn. (\ref{eq:cantedinstab}) diverges at $\epsilon \rightarrow -2$,
the prefactor in front of $(\alpha - \gamma)^2$ in
Eqn. (\ref{eq:cantedinstab})   is negative at small $x$.
% and in low
%fields, $H \stackrel{<}{\sim} \mid E_0 \mid\,,\,\,\, 1-(E_0/8 J_{AF})^2$.
 At $H \rightarrow 0$, this coefficient changes sign at $ x \approx
0.215$ (cf. Appendix \ref{app:FMinstab}, and Eqn.(\ref{eq:FMstable})).

\section{THE UNDERSATURATED FERROMAGNETIC STATE AT LOW $T$}
\label{app:FMinstab}

In this Appendix, we present  results on the breakdown of the
effective exchange approximation and on the low-temperature stability
of the undersaturated ferromagnet.

At $H=0$, the first term in the expansion of $\delta F_1$ (see
Eqn. (\ref{eq:domega1fm})) in powers of $\delta M = \cos \alpha_1 -M$,
\begin{equation}
\delta F_1 (M,T)= A(M,T) \delta M + B(M,T) (\delta M)^2 + ...\,\,,
\label{eq:expomega}
\end{equation}
is proportional to the effective exchange constant, $A=4J_{eff}^{FM}(M)$.
If the temperature is not too low, this linear term (which generates
the effective  exchange approximation) provides a
qualitatively reasonable approximation for $\delta F_1$ 
(see Fig. \ref{fig:DEFM}). Thus, the system
behaves as a Heisenberg ferromagnet with an $M$-dependent exchange
constant. As explained in Sect. \ref{sec:mfe}, 
for sufficiently large
values of $J_{AF}$ (see Eqn. (\ref{eq:intermediate})), $J_{eff}^{FM}$
decreases with decreasing $T$ so that $|J_{eff}^{FM}(M(T))|
\stackrel{<}{\sim}T$. 
%Thus, the effective exchange magnet resembles a Heisenberg 
%ferromagnet in the {\em high-temperature} regime of $T
%\sim T_C$. 
Within the effective exchange approximation, $M_0-M(T)
\propto T$ at $T \rightarrow 0$.

The effective exchange approximation, however, breaks down at low $T$,
when the second term on the r.\ h.\ s. of Eqn. (\ref{eq:expomega})
becomes dominant. This situation (which is depicted in
Fig. \ref{fig:freezepolar}) is due to the fact that the coefficient $B$,
\begin{equation}
B(M,T) \approx \frac{M^2}{4 t^3} 
\left\{ \mid E_0 \mid -
\int_{-2}^{\mu_0} \epsilon^2 J_0(\epsilon)  \nu_0(\epsilon) d
\epsilon \right\}
\label{eq:coeffb} 
\end{equation} 
does not vanish at $M \rightarrow M_0$.
In Eqn. (\ref{eq:coeffb}), $t$ is given by Eqn. (\ref{eq:btfm}), and
\begin{equation}
J_0(\epsilon)= \int_{-2}^{2} {\cal P}\,\frac{\nu_0(\epsilon)}{\eta
- \epsilon} d \eta\,, 
\label{eq:defJ0}
\end{equation}
$\mu_0$ and $\nu_0(\epsilon)$ are the chemical potential and the
density of states in the unrenormalized  ($t=1$) band.

At $M_0-M(T) \ll \sqrt{T}$, the linear in $\delta M$
term in Eqn.  (\ref{eq:expomega}) can be omitted altogether.
We then find that $\delta M =0$ corresponds to an energy minimum if
\begin{equation}
\mid E_0 \mid \,> \int_{-2}^{\mu_0} \epsilon^2 J_0(\epsilon)
\nu_0(\epsilon) d \epsilon\,.
\label{eq:FMstable}
\end{equation}
In this case, the fluctuations of $\cos \alpha$ at low $T$ are confined to the
vicinity of $M_0$ (see
Fig. \ref{fig:freezepolar}). 

When the inequality (\ref{eq:FMstable}) is not satisfied, the
undersaturated FM phase is expected to become unstable at low $T$.
It is easy to see that in any dimensionality $D>1$, the inequality 
(\ref{eq:FMstable}) is violated at $x \rightarrow 0$. This follows from
the fact that,
when $\epsilon$ approaches the bottom of the band, 
\begin{equation}
-\epsilon J_0(\epsilon) = 1- \int_{-D}^{D} {\cal P}\frac{\eta \nu_0(\eta)}
{\eta -\epsilon}
d\eta > 1\,.
\end{equation}
On the other hand, in 2D or in higher dimensions, the inequality
(\ref{eq:FMstable}) is always satisfied for sufficiently large $x$. It is
easy to see that the ratio of the l.\ h.\ s. of  (\ref{eq:FMstable})
to the r.\ h.\ s. increases as the maximum of $\nu_0(\epsilon)$ at
$\epsilon = 0$ becomes more pronounced. Let us consider the extreme
case of a constant density of states, $\nu_0(\epsilon) \equiv 1/4$, and
calculate both sides of  (\ref{eq:FMstable}) at $x=0.5$. We find:
\[ \mid E_0 \mid = \frac{1}{2}\,,\,\,\,\,\,\,
\int_{-2}^0 \epsilon^2 J_0(\epsilon) \nu_0(\epsilon) d \epsilon =
\frac{1}{2} \left( \frac{2}{3} {\rm \, ln}2 + \frac{1}{3} \right)\,,\]
so that the condition  (\ref{eq:FMstable}) indeed is valid. 
Numerical calculations show that in 2D, inequality (\ref{eq:FMstable})
holds for $x > x_c \approx 0.215$. We anticipate that the value of $x_c$
in 3D is lower. We also expect that, similarly to the ferromagnetic or
canted state 
at $T=0$ (see Appendix \ref{app:zerot}), the undersaturated FM state at
low $T$ may become metastable at values of $x$ slightly above $x_c$.

We note that the similar stability conditions for the
antiferromagnetic and canted (at small $H$) phases also take the form
of Eqn.
(\ref{eq:FMstable}). These should be distinguished from the weaker
thermodynamic stability condition, $d \mu/d x >0$, mentioned in Section
\ref{sec:phase}. The latter condition (in the antiferromagnetic, canted,
and undersaturated ferromagnetic phases at $H,T \rightarrow 0$) can be
re-written as
\begin{eqnarray}
\frac{d}{dx}(\mu_0 t)&=&\frac{1}{8J_{AF}}\frac{d}{dx}(\mu_0 |E_0|)= 
\nonumber \\
&=&\frac{1}{8J_{AF}}\left\{\frac{|E_0|}{\nu_0(\mu_0)}-\mu_0^2\right\} >0\,,
\label{eq:TDstab}
\end{eqnarray}
and in 2D holds at $x>0.165$. In writing Eqn. (\ref{eq:TDstab}), we
assumed that $M \rightarrow M_0$ at $T \rightarrow 0$; note that this
may be incorrect whenever the inequality (\ref{eq:FMstable}) is violated.

\begin{figure}
\caption{Single-spin fluctuation in the ferromagnetic phase. The bold 
arrow
represents the average magnetization, and the dashed lines 
correspond 
to the
hopping amplitude $b$, which differs from the background hopping 
value 
$t$
(solid lines).}
\label{fig:canted}
\end{figure}

\begin{figure}
\caption{Magnetization vs. temperature in the ferromagnetic phase 
at $H=0$,
$x=0.4$, and  $J_{AF}=0.06$. The
solid, dashed, and dotted lines correspond to the 2D DE--
superexchange
magnet, effective exchange approximation, and usual Heisenberg
ferromagnet, respectively.}
\label{fig:DEFM}
\end{figure}

\begin{figure}
\caption{The behaviour of the sublattice (solid line) and net (dashed 
line)
magnetizations in the canted state at $H=0.01 $, $x=0.4$, and
$J_{AF}=0.06$, in comparison with 
the magnetization of the ferromagnetic state (dotted line).
The dashed-dotted line represents the results for the canting angle,
$\gamma$.}  
\label{fig:Mcanted}
\end{figure}

\begin{figure}
\caption{Mean field results for the case of strong superexchange,
$J_{AF}=0.08$, $x=0.3$. The solid and dashed-dotted lines represent the result 
for the
sublattice magnetization $m$ and canting angle $\gamma$ for $H=0.01$. 
The dotted line corresponds 
to the sublattice magnetization $m_{AFM}$ of N\'{e}el AFM phase at $H=0$.} 
\label{fig:flop}
\end{figure}

\begin{figure}
\caption{Phase diagrams of the DE--superexchange magnet for
$J_{AF}=0.06$  at $H=0$ (a) and $H=0.01$ (b), and for $J_{AF}=0.05$
at  $H=0$ (c) and  $H=0.01$ (d),
showing the ferro-, antiferro- (flop-phase at $H>0$), paramagnetic,
and canted phases (FM, AFM, PM, and CM, respectively). The 
dashed line in Fig. \ref{fig:phase} (a,c) denotes the
boundary of the thermodynamically unstable ($d \mu /d x <0$) region,
and $x=x_1$ in  Fig. \ref{fig:phase} (c,d) separates undersaturated
($x<x_1$) and saturated low-temperature regimes at $H=0$.  The
behaviour of the system is symmetric with respect to quarter-filling, 
$x=0.5$.}
\label{fig:phase}
\end{figure}

\begin{figure}
\caption{Single-spin perturbation energy, $\delta F^{(0)}(\alpha)$, in
the ferromagnetic state at $T=0$ in zero field, for $x= 
0.4$ and $J_{AF}=0.04$. Solid line represents the exact result (see
Eqn. (\ref{eq:dF0})), while the
dashed line corresponds to the effective exchange approximation,
Eqn. (\ref{eq:eeaT0}). The dotted
line represents the exact result for $\delta
F^{(0)}(\alpha)$ at $x=0.15$, $J_{AF}=0.025$.} 
\label{fig:domegafm}
\end{figure}

\begin{figure}
\caption{The function $\delta F^{(0)}(\alpha)$ 
in the canted state at $T=0$. The left panel corresponds to
$x=0.4$, $J_{eff}=0.08$, $H=0.01$. The dashed line represents the band
contribution. In the right panel, $\delta F^{(0)}(\alpha)$ is plotted 
for $J_{eff}=0.06$ and $H=0.01$ at $x=0.4$ (solid line), $x=0.25$
(dashed line) and $x=0.15$ (dotted line).}
\label{fig:domegacanted}
\end{figure}

\begin{figure}
\caption{The energy cost, $\delta F_1$, of a single-site
fluctuation in the ferromagnetic case at low $T$ (see Appendix
\ref{app:FMinstab}). The pronounced
minimum of $\delta \Omega$ at the average value of the polar angle
$\alpha$ represents a sharp difference from the effective exchange
approximation (dashed line) and causes the fluctuations of $\alpha$
(but not of the azimuthal angle, $\beta$) to freeze out at low $T$. The plot
corresponds to $H=0$, $x=0.4$, $T=0.002$, and $J_{AF}=0.06$.}
\label{fig:freezepolar}
\end{figure}


\begin{references}
\bibitem{undoped} In this model (which was chosen to fit the known magnetic 
behaviour of the 
undoped perovskite ${\rm La Mn O}_3$), 
ferromagnetic intra-layer exchange results
in ferromagnetic ordering within the lattice layers.
\bibitem{virtual} Our choice of virtual crystal parametrization does
not include but could 
be generalized to incorporate a
$(\pi,0,0)$-type AF ordering.
\bibitem{finiteT} The small ($\sim T/(\mu+2t)$) thermal correction to the
total band energy, originating from the smearing of the Fermi
distribution function, is omitted for simplicity in the text of the
present paper. 
It was, however, taken into account in our numerical calculations
(see Figs. \ref{fig:DEFM}--\ref{fig:phase}).
\bibitem{H0} Eqn. (\ref{eq:intermediate}) refers to the $H=0$ case. At $H>0$,
either ferromagnetic or paramagnetic (with $M \sim H$) solution to the
mean field equations formally exists at all values of $E_0$. 
As $M$ increases from $0+$ to
$1$, the quantity $J_{eff}^{FM}(M)-H/(4M)$ changes sign if $8J_{AF}>
|E_0|+2H$; the latter condition replaces Eqn. (\ref{eq:intermediate}).
\bibitem{fieldM0} In the presence of a magnetic field, $M_0$ is given by
the root of equation $4 M J^{FM}_{eff}(M)=H$. Thus, the noticeable
increase in $M_0$ requires very high fields, $H \sim J_{AF} \sim t_0 x$.
\bibitem{integral} In evaluating the average values in
Eqns. (\ref{eq:mfec1}--\ref{eq:mfec2}), the integrations over the
azimuthal angle $\beta_1$ can be performed analytically.
\bibitem{susce} If the transition temperature $T_1$ is sufficiently low,
so that $M_0-M(T_1) \ll T_1$, the associated susceptibility jump
is given by $\chi_{canted}-\chi_{ferro} \approx 1/[4 T_1 (1-M_0^2)]$.
\bibitem{Chubukov} While it is not
clear whether  undersaturation is present in the case of quantum
spins, we note that
this would be not inconsistent with the results obtained for the 
somewhat
similar case of quantum ferrimagnets (see: A. V. Chubukov,
K. I. Ivanova, P. Ch. Ivanov, and E. R. Korutcheva, J. Phys.: 
Cond. Matt. {\bf 3}, 2665 (1991)).
\bibitem{flip}The spin-flip transition between the flop-phase and a paramagnet 
occurs at temperature $T_1$ which is again given by Eqn. (\ref{eq:T1}), 
$T_1 \rightarrow T_N$ at $H \rightarrow 0$  (note that
in an antiferromagnet, $M \propto H$ in weak fields).
\bibitem{difference} One should distinguish between Rietveld
measurements\cite{Rietveld}  and
SQUID magnetometer results 
(Refs. \onlinecite{Raveau,Moritomo}), which can be affected by
possible interlayer canting.
\bibitem{dos} We make use of the facts that
$\nu(\epsilon)=\nu(-\epsilon)$ and $\int_{-2t}^{2t}\nu(\epsilon)
d \epsilon = 1$.
\end{references}
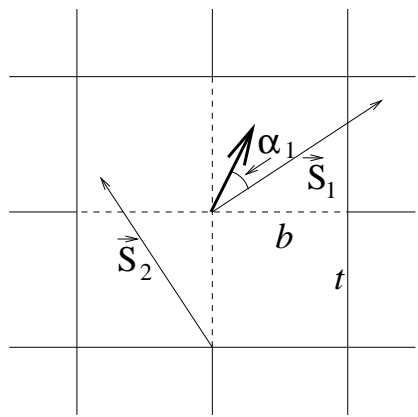
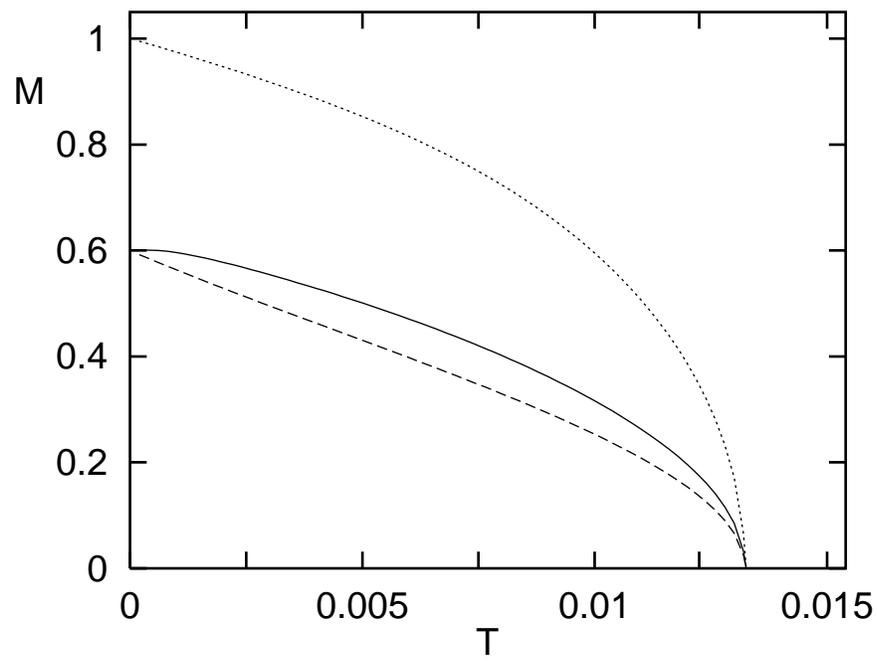
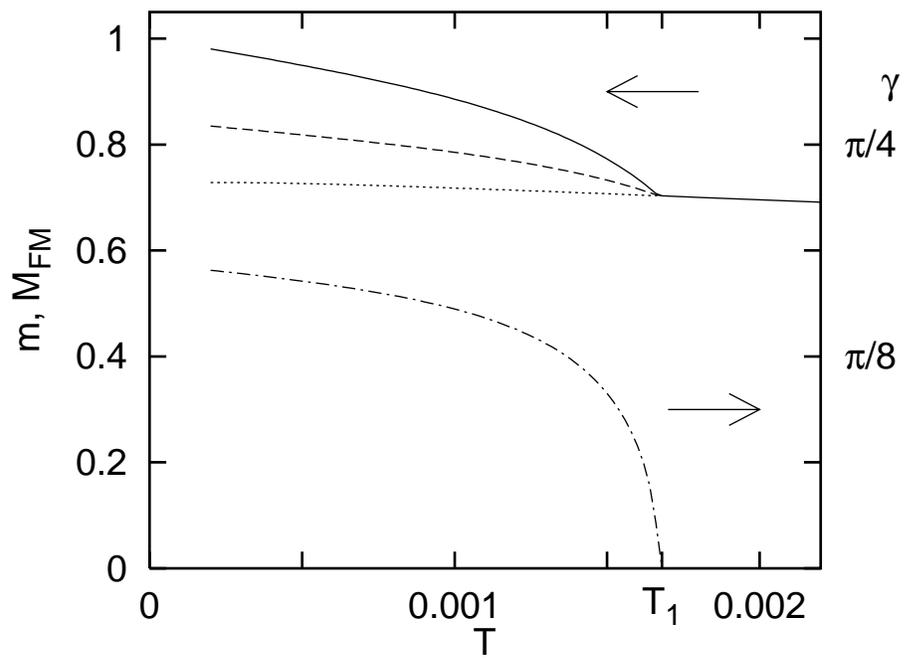
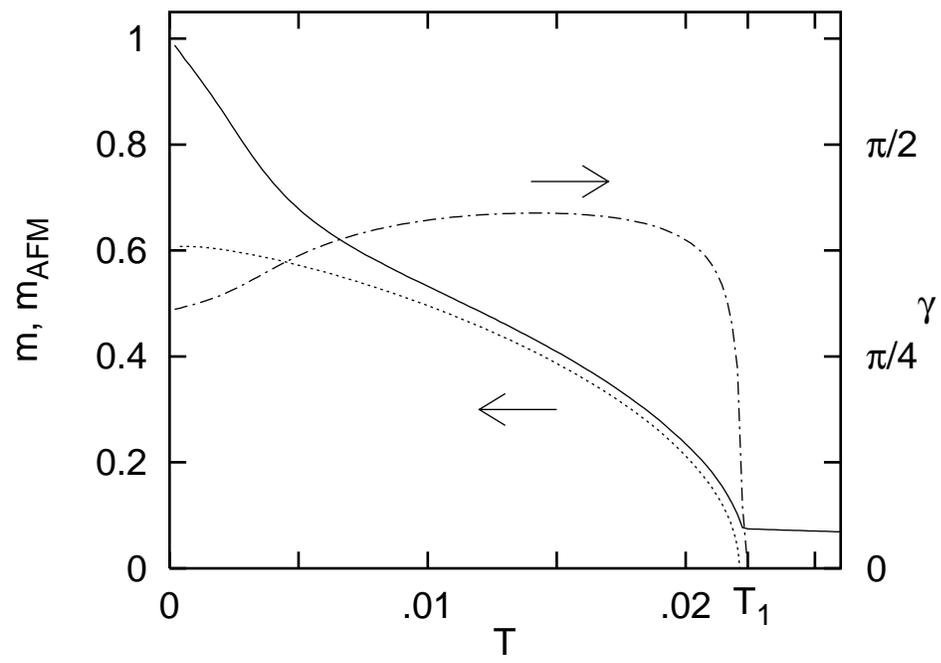
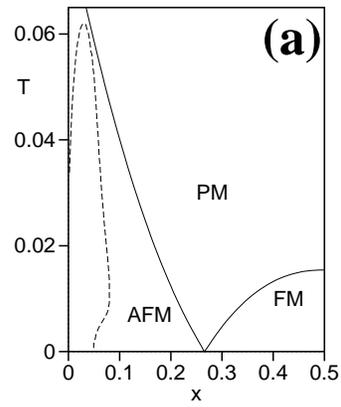
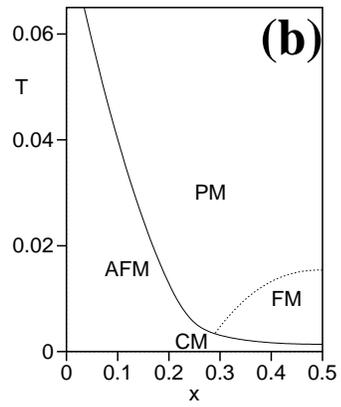
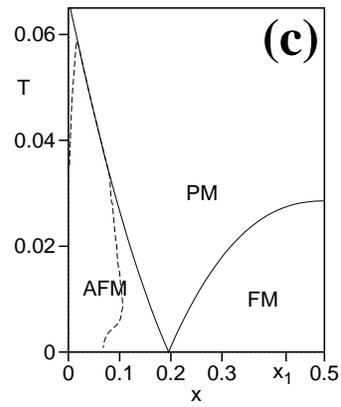
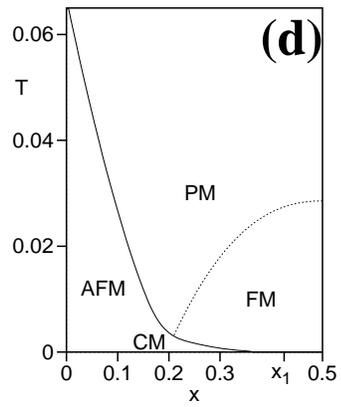
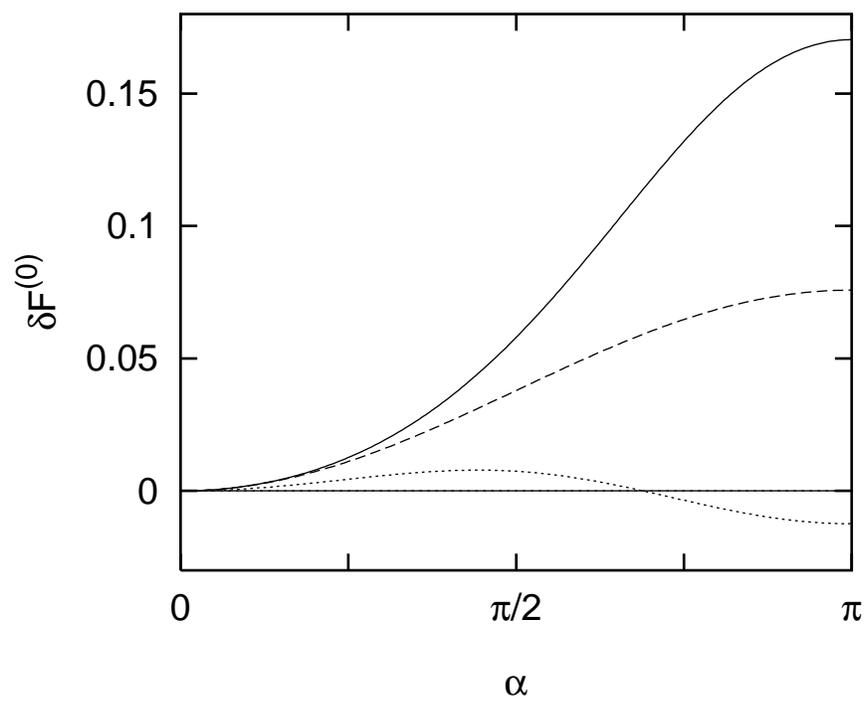
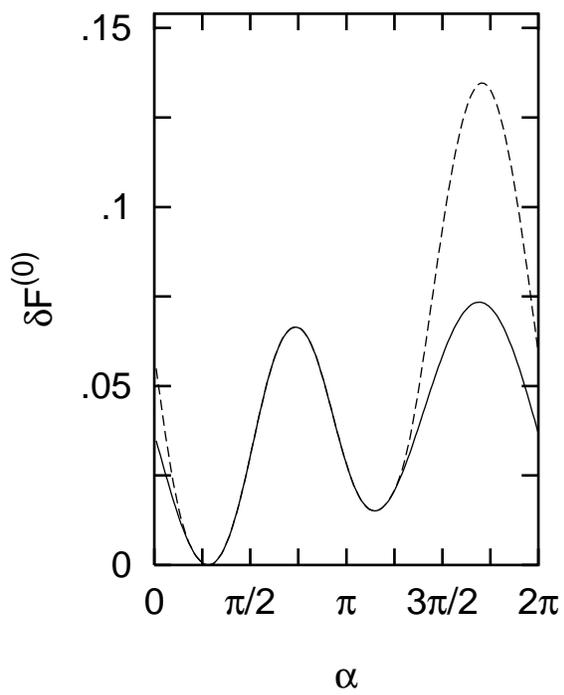
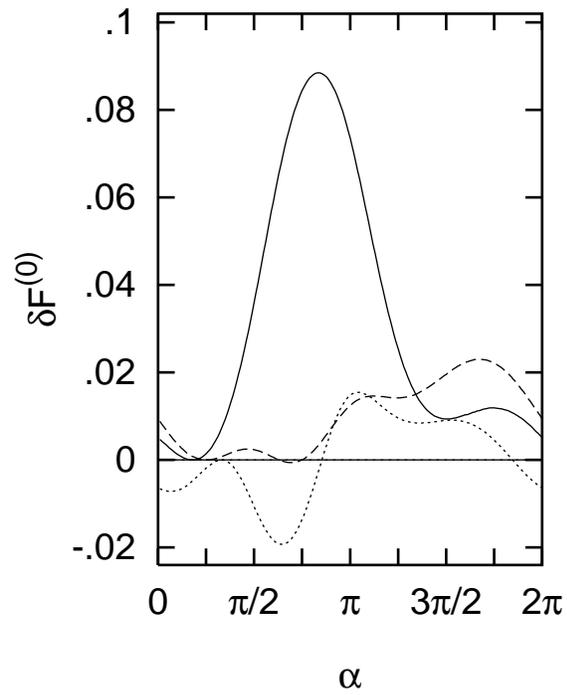
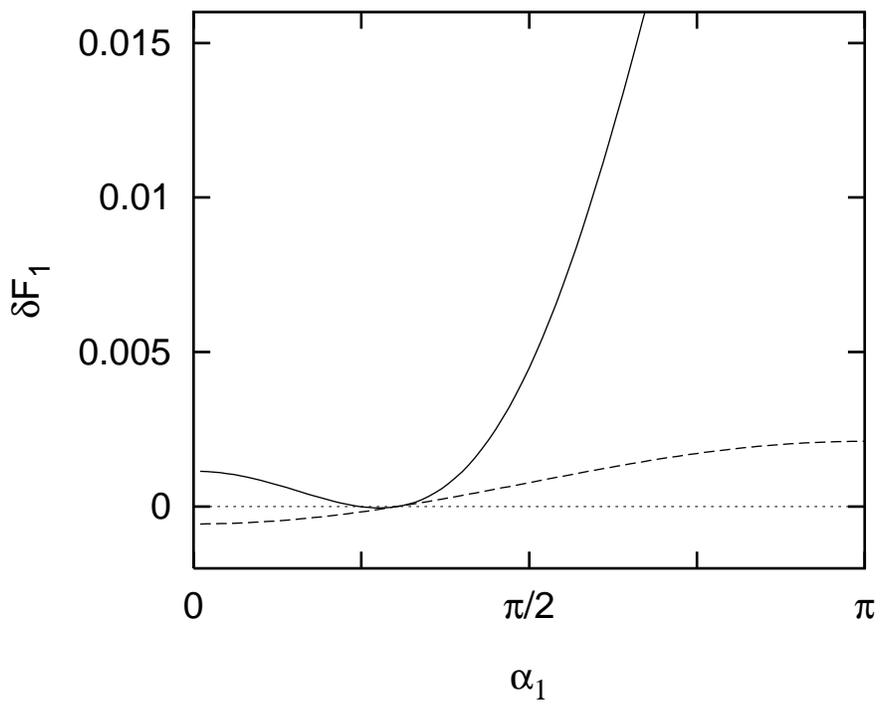
\end{document}